\documentclass[11pt,preprint]{aastex}

\begin{document}
\title{On the Benefits of Promoting Diversity of Ideas}

\author{Abraham Loeb\\Institute for Theory \& Computation\\
Harvard University\\60 Garden St., Cambridge, MA 02138}

\bigskip
\bigskip
\bigskip

{\it ~\\``It ain't what you don't know that gets you into
trouble. \\It's what you know for sure that just ain't so. ''}

\noindent
{\it Mark Twain}

A very common flaw of astronomers is to believe that they know the
truth even when data is scarce.\footnote{This essay was written as the
author was preparing a lecture on a newly discovered population of
``Fast Radio Bursts'', which is widely assumed to originate at
cosmological distances based on limited evidence. The video of the
lecture is available at {\it
https://www.youtube.com/watch?v$=$aHVx6FCHsCg}} This fault is the
trademark of a data-starved science. It occasionally leads to major
blunders by which the scientific community makes the wrong strategic
decision in its research plans, causing unnecessary delays in finding
the truth.

Let me illustrate this phenomenon with ten examples, in chronological order.

\begin{itemize}

\item {\bf Large telescopes.}  In 1909, Edward Charles Pickering, who
served as director of the Harvard College Observatory from 1877 until
1919, argued that telescopes had reached their optimal size and there
was no advantage gained in seeking larger apertures. In December 1908
he wrote\footnote{{\it
http://www.gutenberg.org/files/15636/15636-h/15636-h.htm}} in an
article titled ``The Future of Astronomy'': "... It is more than
doubtful, however, whether a further increase in size is a great
advantage. Much more depends on other conditions, especially those of
climate, the kind of work to be done and, more than all, the man
behind the gun. The case is not unlike that of a battleship. Would a
ship a thousand feet long always sink one of five hundred feet? It
seems as if we had nearly reached the limit of size of telescopes, and
as if we must hope for the next improvement in some other direction."

Pickering's blunder led to a major blow for observational astronomy in
the east coast relative to the west coast of the US.  Just before his
article was published, George Ellery Hale obtained first light on the 60
inch telescope at Mt Wilson Observatory, which became one of the most
productive telescopes in astronomical history. Around the same time,
Hale received funding from Hooker and Carnegie to create a larger
telescope. In 1917 the 100 inch telescope was completed; Hubble and
Humason used it to discover the expansion of the universe. The 100
inch telescope was surpassed by the 200 inch telescope at Mt Palomar
in 1948, which played a key role in the discovery of radio galaxies
and quasars and in studies of the intergalactic medium.  Clearly,
bigger telescopes continued to benefit astronomy as technology
improved.

\item {\bf Composition of the Sun.}  During her PhD thesis in 1925
(which was the first PhD in Astronomy at Harvard-Radcliffe), Cecilia
Payne-Gaposchkin interpreted the solar spectrum based on the Saha
equation and concluded that the Sun's atmosphere is made mostly of
hydrogen. While reviewing her dissertation, the distinguished
Princeton astronomer Henry Norris Russell convinced her to avoid
the conclusion that the composition of the Sun is different from that
of the Earth, as it contradicted the conventional wisdom at the time.

\item {\bf Maser and complex molecules.} When Charlie Townes worked on
his experimental demonstration of the maser in 1954, two Nobel
laureates, Isidor Isaac Rabi and Polykarp Kusch, tried to stop him by
saying\footnote{{\it Edward Teller Lectures: Laser and Inertial Fusion
Energy}, Eds. H. Hora \& G. Miley, Imperial College Press (2005),
page 4.}: ``Look, you should stop the work you are doing. It isn't
going to work. You know it's not going to work, we know it's not going
to work. You're wasting money. Just stop!''. Three months later, the
maser worked. Similar circumstances repeated when Townes was
determined to discover complex molecules in space and was resisted by
astronomers who argued that the interstellar gas density is so low and
the UV illumination so intense that any surviving molecules would be
too scarce to be detectable\footnote{{\it Revealing the Molecular
Universe: One Antenna is Never Enough},  Eds. D. Backer,
J. Moran, \& J. Turner, ASPC, {\bf 356}, 81 (2006).}.

\item {\bf X-ray Astronomy.} In the early 1960s, a panel of experts
was assembled by NASA to evaluate the merit of a proposal to launch an
X-ray telescope into space (since the Earth's atmosphere blocks
X-rays). The panel concluded that the scientific motivation for an
X-ray telescope was weak, since most of the X-ray sources would be
flaring stars. The launch of an X-ray telescope by NASA was therefore
delayed by half a decade, after which astronomers discovered X-ray
emission from numerous other sources, such as accreting black holes
and neutron stars, supernova remnants, and galaxy clusters.

\item {\bf Dark matter.} In the early 1970s Jerry Ostriker gave a talk
at Caltech describing the case, developed by him in collaboration with
Jim Peebles and Amos Yahil, for spiral galaxies having dark matter
halos that comprises most of their mass. The observers in the
audience were contemptuous of the idea and dismissed it as a wild
theoretical speculation.

\item {\bf Gravitational lensing.} Around 1980, shortly after the
discovery of the first gravitational lens 0957$+$561A,B, Ed Turner
at Princeton was advised by a highly distinguished astronomer
not to spend much time working on gravitational lenses because they
will turn out to be ``useless curiosities''. For a few years, lenses
were widely regarded by astronomers as unimportant and it was almost
impossible to get observing time or grants to study them.

\item {\bf Cosmology.} Around 1990, during my term as a postdoctoral
fellow at Princeton, I asked a prominent astronomer from another
prestigious academic institution whether they would consider a junior
faculty hire in the field of theoretical cosmology. He replied: ``we
might contemplate this possibility if we could only convince ourselves
that cosmology is a science.'' Two years later, in 1992, the COBE
satellite reported the detection of microwave background anisotropies.

\item {\bf High-redshift galaxies.} Piero Madau at UCSC once told me
that a paper he wrote in the mid 1990s on intergalactic absorption and
the colors of high redshift galaxies had great difficulties getting
published because the referee kept arguing: ``we all know that there
are no normal galaxies above a redshift of two''.
 
\item {\bf Kuiper Belt Objects.} David Jewitt at UCLA could not get
telescope time nor funding for attempts to detect the conjectured
population of Kuiper Belt Objects. He used observing time and funding
he received for other projects, until he finally discovered the first
of these objects in the outer solar system with Jane Luu in 1992, using
the 88 inch telescope at Mauna Kea.

\item {\bf Close-in Jupiters.} The first planets ever discovered
around a main sequence star other than the Sun had masses similar to
Jupiter but were orders of magnitude closer to their host star than
Jupiter is to the Sun. This can be simply understood as a selection
effect, since the reflex motion of a star due to a close-in planet is
much easier to detect than the motion induced by a distant planet. But
because Jupiter is considerably farther out from the center of the
solar system, time allocation committees on major telescopes declined
proposals to search for close-in Jupiters for years based on the
argument that such systems would deviate dramatically from the
architecture of the solar system and hence are unlikely to
exist. Theoretical prejudice prevented HD 114762b,
discovered\footnote{Latham, D.W., Stefanik, R. P., Mazeh, T., Mayor,
M., \& Burki, G., Nature, {\bf 339}, 38 (1989).} by Latham et al. in
1989, from being recognized as a planet for over 6 years, until after
Mayor \& Queloz announced\footnote{Mayor, M., \& Queloz, D., Nature,
{\bf 378}, 355 (1995).} the discovery of 51 Peg b in 1995 and Butler
\& Marcy found similar examples.  As it turns out, Otto Struve has
already suggested\footnote{Struve, O., The Observatory, {\bf 72}, 199
(1952).}  in 1952 that close-in planets may exist and would be easy to
find through both radial velocity and transit observations, but his
paper was completely ignored because of theoretical priors.

\end{itemize}

These examples and many more like them (starting with the ancient view
that the Earth is at the center of the Universe and the Sun revolves
around it), demonstrate that progress in astronomy can be delayed by
the erroneous proposition that we know the truth even without
experimental feedback.  Lapses of this type can be avoided by a honest
and open-minded approach to scientific exploration, which I label as
having a ``non-informative prior'' (so called ``Jeffreys prior'' in
Bayesian statistics). This unbiased approach, which is common among
successful crime detectives, gives priority to evidence over
imagination, and allows nature itself to guide us to the correct
answer. Its basic premise is humility, the recognition that nature is
much richer than our imagination is able to anticipate.

Uniformity of opinions is sterile; the co-existence of multiple ideas
cultivates competition and progress.  Of course, it is difficult to
know in advance which exploratory path will bear fruit, and the back
yard of astronomy is full of novel ideas that were proven wrong. But
in order to make the discovery process more efficient, telescope-time
allocation committees and funding agencies should dedicate a fixed
fraction of their resources (say 10-20\%) to risky explorations. This
can be regarded as affirmative action to promote diversity of ideas,
which is as important for the progress of science as the promotion of
gender and ethnic diversity.

\bigskip
\bigskip
\bigskip

\acknowledgements I thank Dani Maoz, Robb Scholten, Amiel Sternberg,
Scott Tremaine, Ed Turner and Josh Winn for helpful comments on the
manuscript.

\end{document}